\documentstyle[prb,aps]{revtex}
\begin{document}
\draft
\title{
Superconducting, Ferromagnetic and Antiferromagnetic Phases
in the $t-t'$ Hubbard Model}
\author{J. V. Alvarez $^1$,
 J. Gonz\'alez $^2$, F. Guinea $^3$ and M. A. H. Vozmediano $^1$ \\}
\address{
        $^1$Departamento de Matem\'aticas.
        Universidad Carlos III.
        Butarque 15.
        Legan\'es. 28913 Madrid. Spain. \\
        $^2$Instituto de Estructura de la Materia.
        Consejo Superior de Investigaciones Cient{\'\i}ficas.
        Serrano 123, 28006 Madrid. Spain. \\
        $^3$Instituto de Ciencia de Materiales.
        Consejo Superior de Investigaciones Cient{\'\i}ficas.
        Cantoblanco. 28049 Madrid. Spain.}
\date{\today}
\maketitle
\begin{abstract}
We apply a renormalization group approach to the determination
of the phase diagram of the $t-t'$ Hubbard model at the Van Hove filling,
as function of $t' / t$, for small values of $U / t$.
The model presents ferromagnetic, antiferromagnetic and d-wave
superconducting phases. Antiferromagnetism and d-wave
superconductivity arise from the same interactions, and compete
in the same region of parameter space.
\end{abstract}

In recent years great effort has been devoted to study the role
of Van Hove singularities in two-dimensional electron
liquids\cite{hove,schulz,dzya,lee,mark,tsuei,vanhove,ioffe}.
Most part of the interest stems from the evidence, gathered from
photoemission experiments, that the hole-doped copper oxide
superconductors tend to develop very flat bands near the Fermi
level\cite{photo,gofron}.
The model most widely used for the
Cu-O planes is the Hubbard model,
which also shows
strong antiferromagnetic correlations. The
correspondence with the phenomenology of the high-T$_c$
materials is supposed to be attained in the intermediate to
strong coupling regime. Weak coupling approaches
to the model have shown that it is more likely to develop a
spin-density-wave instability than superconductivity near
half-filling\cite{schulz,dzya,zanchi}.

The inspection of the actual Fermi line of most of the copper oxide
superconductors shows, however, a sensible departure from
nesting. In the context of models with on-site interaction, the
dispersion relation seems to be best fitted by introducing both
nearest neighbor $t$ and next-to-nearest neighbor $t'$,
accomodated in the so-called $t-t'$ Hubbard model\cite{ttp}.
In this model, the absence of perfect nesting implies that
the antiferromagnetic correlations have a less divergent
behavior when the Fermi level lies at the Van Hove singularity.
For this reason, it is best suited for the implementation of a
renormalization group (RG) approach to the ground state properties,
that is needed anyhow to deal with the logarithmic singularities
arising from the divergent density of states\cite{hirsch}.

In this Letter we look for the instabilities of the $t-t'$
Hubbard model, filled up to the level of the Van Hove
singularity, following the wilsonian RG program of Refs.
\onlinecite{shankar,pol}. Our approach is that of integrating
virtual states of two energy slices above and below the Fermi
level in an energy range given by the cutoff $E_c$,
$E_c - |d E_c| < |E| < E_c$. We are interested in the scaling
behavior of the interactions and correlations under a
progressive reduction of the cutoff, which leads to the
description of the low-energy physics about the Fermi level.

In the RG process, we have to make sure first that the
interaction does not scale to zero at the classical level,
that is, it displays marginal
behavior as we approach the Fermi line. When the cutoff is
sufficiently close to the Fermi level, as shown in Fig.
\ref{one}, most part of the states at hand are in the
neighborhood of any of the two Van Hove points $(\pi ,0)$ $(A)$
and $(0, \pi)$ $(B)$, and the dispersion relation may be
approximated by two respective patches about each of them:
\begin{equation}
\varepsilon_{A,B} ( {\bf k} ) \approx \mp ( t \mp 2 t' ) k_x^2 a^2
\pm ( t \pm 2 t' ) k_y^2 a^2
\end{equation}
where $a$ is the lattice constant. The angle between the 
two separatrices is $2 \arctan  [ ( t + 2 t' ) / ( t - 2 t' ) ]$.
The low-energy effective action may be written in the form
\begin{eqnarray}
S  & = & \int d \omega d^2 k \sum_{\alpha,\sigma} \left( \omega
\; a^{+}_{\alpha,\sigma}({\bf k}, \omega )
   a_{\alpha,\sigma}({\bf k}, \omega )
- \varepsilon_{\alpha} ( {\bf k}) \;
   a^{+}_{\alpha,\sigma}({\bf k}, \omega )
   a_{\alpha,\sigma}({\bf k}, \omega ) \right)   \nonumber   \\
  &   &  - U \int d\omega d^2 k \;
 \rho_{\uparrow } ({\bf k}, \omega) \;
      \rho_{\downarrow } (-{\bf k}, -\omega)
\label{actk}
\end{eqnarray}
where $a_{\alpha,\sigma} (a^{+}_{\alpha,\sigma})$ is an electron
annihilation (creation) operator ($\alpha$ labels the Van Hove point),
$ \rho_{\uparrow ,\downarrow }$ are the density operators in
momentum space, and the momentum integrals are restricted to
modes within the energy cutoff, $|\varepsilon_{\alpha} \; ({\bf k})|
\leq E_c $. It is a fact that the action (\ref{actk}) defines a
model that is scale invariant at the classical level, that is,
the different terms that build up (\ref{actk}) are invariant
altogether under a rescaling of the cutoff $E_c$\cite{vanhove}.
This implies that the interaction in
the $t-t'$ Hubbard model has marginal behavior, at least at the
classical level, and it is therefore susceptible of a low-energy
description within the RG framework.

In our low-energy model, the two patches for the
dispersion relation give rise
to an additional flavor index. The number of possible interactions
proliferates accordingly.
The different terms are
depicted in Fig. \ref{two} . It might seem that the analysis of
the quantum corrections in the model should bear a great
similarity with the g-ology description of one-dimensional
models. 
In two dimensions, however, kinematical constraints play
a crucial role.
Performing
the wilsonian RG analysis, no singular cutoff dependences are
found in the particle-particle channel, for instance, unless the
two colliding particles have opposite momenta. This conclusion
may be drawn from the general discussion carried out in Ref.
\onlinecite{shankar}. The BCS instability is to be analyzed,
then, as a singularity in a response function computed at a
definite value of the momentum. On the other hand, singular
dependences on the cutoff are found in the
particle-hole channel, when the momentum transfer is about zero
or about the momentum connecting the two Van Hove points, ${\bf Q}
\equiv (\pi , \pi)$.

The interactions shown in Fig. \ref{two} are renormalized by the
high-energy modes at the quantum level. In all the four
cases, the only contribution to second order in perturbation
theory which is $ \sim dE_c /E_c$ is given by a diagram
of the type shown in Fig. \ref{three}. As mentioned before,
the contributions in the particle-particle channel are $O(
(dE_c)^2 )$ for generic values of the incoming and outgoing momenta,
and there are no more diagrams which can be built up from the
interaction in (\ref{actk}). 
To second order we get the RG flow equations

\begin{eqnarray}
E_c \frac{\partial U_{intras}}{\partial E_c}  & = &
 \frac{1}{2\pi^2 t} c \left( U_{intras}^2 + U_{back}^{2}
       \right)    \label{flow1}       \\
E_c \frac{\partial U_{back}}{\partial E_c}  & = &
 \frac{1}{\pi^2 t} c \left( U_{intras}  U_{back} \right)\\
E_c \frac{\partial U_{inters}}{\partial E_c}  & = &
 \frac{1}{2\pi^2 t} c' \left( U_{inters}^2 + U_{umk}^{2} \right)\\
E_c \frac{\partial U_{umk}}{\partial E_c}  & = &
 \frac{1}{\pi^2 t} c' \left( U_{inters} U_{umk}
        \right)      \label{flow4}
\end{eqnarray}
where $c \equiv 1/\sqrt{1 - 4(t'/t)^2}$ and $c' \equiv
\log \left[ \left(1 + \sqrt{1 - 4(t'/t)^2} \right)/(2t'/t) \right]$
are the prefactors of the polarizabilities at zero and ${\bf Q}$
momentum transfer, respectively.

The RG equations
(\ref{flow1})-(\ref{flow4}) describe a flow that drives the
couplings to large values, as the cutoff is sent to the Fermi
line. This is due to the localization in space of the
interactions, which makes them strongly spin-dependent.
The RG flow of extended, spin-independent, interactions is
towards lower couplings, as discussed in Ref. \onlinecite{vanhove}. 
In that case,
RPA-like screening is the dominant effect. 
The limit of a short range interaction is
special in that there is a cancellation between the 
`particle-hole bubble' and `vertex-correction' second order diagrams. For a
purely local interaction like that of the $t-t'$ Hubbard model,
the only contribution left correspons to the antiscreening
diagram shown in Fig. \ref{three}.

In order to apply the above RG equations to the $t-t'$ model, we
compute the flow starting with all the couplings set to the
original value $U$ of the on-site interaction. Under this
initial condition, it is clear that $U_{intras} = U_{back}$ and
$U_{inters} = U_{umk}$, all along the flow. The renormalized
vertices show
divergences at certain values of the frequency, measured in
units of the cutoff $E_c$. We have
\begin{eqnarray}
U_{intras}  & = & \frac{U}{1 + U \: c/( \pi^2 t) \:  \log(\omega
       /E_c) }     \label{intra}       \\
U_{inters}  & = & \frac{U}{1 + U \: c'/( \pi^2 t) \:  \log(\omega
       /E_c) }      \label{inter}
\end{eqnarray}
We interpret the divergences of the vertices in the same way as
in the RPA, as signalling the development of an
ordered phase in the system.

The precise determination of the instability which dominates for
given values of $U$ and $t'$ is accomplished by analyzing the
response functions of the system. The procedure
is similar to that followed in the study of one-dimensional
electron systems\cite{book}. We adopt again a RG
approach to their computation, which takes into account the
scaling behavior of the interactions (\ref{intra}) and
(\ref{inter}). We analyze
ferromagnetic, antiferromagnetic, superconducting and CDW
correlations. We assume that the actual 
instabilities of the model
at low temperatures are some of these.

It is easily seen that the operators related to
charge-density-wave and s-wave superconducting instabilities do
not develop divergent correlations at small $\omega $. The phase
diagram in the $t'-U$ plane is drawn by looking at the
competition among ferromagnetic, antiferromagnetic and d-wave
superconducting instabilities. The ferromagnetic response
function $R_{FM}$, for instance, is given by the correlation of
the uniform magnetization, $\rho_{\uparrow} (0,\omega ) -
\rho_{\downarrow} (0,\omega ) $. The first perturbative terms
for this object are built from a couple of one-loop
particle-hole diagrams linked by the interaction.
Each particle-hole
bubble has a logarithmic dependence on the cutoff $E_c$, with
the prefactor $c = 1/\sqrt{1 - 4(t'/t)^2}$. The iteration
of bubbles can be taken into account by differentiating with
respect to $E_c$ and writing a self-consistent equation for
$R_{FM}$, which turns out to be
\begin{equation}
\frac{\partial R_{FM}}{\partial E_c} = -  \frac{2c}{\pi^2 t}
  \frac{1}{E_c}  +  \frac{c}{ \pi^2 t} \left( U_{intras} +
  U_{inters} \right) \frac{1}{E_c} R_{FM}
\label{ferro}
\end{equation}
The antiferromagnetic response function $R_{AFM}$ can be dealt
with in a similar fashion, by looking at correlations of the
operator $\rho_{\uparrow} ({\bf Q},\omega ) -
\rho_{\downarrow} ({\bf Q},\omega ) $. This leads to the RG
equation
\begin{equation}
\frac{\partial R_{AFM}}{\partial E_c} = -  \frac{2c'}{\pi^2 t}
  \frac{1}{E_c}  +  \frac{c'}{ \pi^2 t} \left( U_{back} +
  U_{umk} \right) \frac{1}{E_c} R_{AFM}
\label{aferro}
\end{equation}
We recall that $U_{intras} +
U_{inters}$ and $U_{back} + U_{umk}$ have the same flow, within
the present model. Therefore, we may discern at once that
whenever $c > c'$ the ferromagnetic response function $R_{FM}$
prevails over $R_{AFM}$.

Finally, there remains the response function
for d-wave superconductivity $R_{SCd}$, which is given by the
correlation of the operator $\sum_{{\bf k}} \left(
a^{+}_{A\uparrow }({\bf k}) a^{+}_{A\downarrow }({\bf -k}) -
a^{+}_{B\uparrow }({\bf k}) a^{+}_{B\downarrow }({\bf -k}) +
 h.c. \right)$. The computation
within the RG framework becomes now a little bit more subtle,
since the diagrams at strictly zero total momentum display a
$\log^2 E_c$ dependence on the cutoff. As the derivative with
respect to $E_c$ is taken, it becomes clear that the logarithmic
dependence left corresponds to the divergent density of states
at the Van Hove singularity. The RG equation for $R_{SCd}$ reads then
\begin{equation}
\frac{\partial R_{SCd}}{\partial E_c} = -  \frac{c}{2\pi^2 t}
  \frac{\log (E_c/\omega) }{E_c}  -
  \frac{c}{2 \pi^2 t} \left( U_{intras} -
  U_{umk} \right) \frac{\log (E_c/\omega) }{E_c} R_{SCd}
\label{scd}
\end{equation}
This equation also shows an homogeneous scaling of $R_{SCd}$ on
$\omega / E_c$, like in the previous cases.

From inspection of Eq. (\ref{scd}), it is clear that divergent
correlations in the d-wave channel arise for $U_{intras} -
  U_{umk} < 0$. According to the above results, this only
happens for $c < c'$, that is, outside the region of the phase
diagram where $R_{FM} > R_{AFM}$. This confirms that a
ferromagnetic regime sets in for values of $t'$ above
the critical value $t_{c}' \approx 0.276 t $ at which $c = c'$. For
values below $t_{c}'$, there is a competition between $R_{AFM}$
and $R_{SCd}$, which requires the analysis of the respective
behaviors close to the critical frequency at which the response
functions diverge. As a general trend, the response function
$R_{SCd}$ dominates over $R_{AFM}$ in the regime of weak
interaction, the strength being measured with regard to both the
bare coupling constant and the value of the $c'$ parameter. The
reason for such behavior is that at weak interaction strength
the RG flow has a longer run to reach the critical frequency,
and at small frequencies the logarithmic density of states in
Eq. (\ref{scd}) makes $R_{SCd}$ to grow larger. The border where
the crossover between the antiferromagnetic and the superconducting
instability takes place is shown in the $t'-U$ phase diagram of
Fig. \ref{four}. At sufficiently large values of $U$ and
small values of $t'$, the leading instability of the system
turns out to be antiferromagnetism. This is in agreement with
weak coupling RG analyses applied to the Hubbard
model\cite{schulz,dzya,zanchi}.

Thus,  there exists a region of the phase diagram
where superconductivity is the leading instability. 
We remark that
this result is obtained within a RG approach that provides a
rigorous computational framework, with no other assumption than
the weakness of the bare interaction. The instabilities are led
by an unstable RG flow, and the singular behavior of the
response functions is interpreted in the same fashion than in a
standard RPA computation. 
The wide range for superconductivity is consistent with the
results from quantum Monte Carlo computations\cite{mc},
as well as with results obtained by exact diagonalization of
small clusters in the strong coupling regime\cite{nd}.

On the other hand, the renormalized
interactions remain most part of the flow in the weak coupling
regime for $U/t < 1$. We observe that the most interesting region
in the phase diagram of Fig. \ref{four} starts near $U/t \approx
1$, where the
critical frequencies are $\omega_c \sim 10^{-2} E_c$. 
This order
of magnitude corresponds to sizeable critical temperatures
($\sim$ 100 K) if we assign to $E_c$ a value of the
order of the conduction bandwidth in the cuprates ($\sim$ 1eV).

The other relevant conclusion
within our RG approach is the existence of a ferromagnetic
regime in the $t-t'$ model, above a certain value of the $t'$
parameter. This is consistent with the results obtained in Ref.
\onlinecite{ferro} close to $t' = 0.5 t$. Though our results refer
to the weak coupling regime, they show that antiferromagnetic,
ferromagnetic and superconducting phases are all realized in the
$t-t'$ Hubbard model. The superconducting instability has
greater strength at the boundary with the antiferromagnetic
instability, as it also happens in other approaches to
high-$T_c$ superconductivity\cite{afm}.

In our case, the diagrams responsible for the apppearance of
superconductivity cannot be interpreted in terms of the exchange
of antiferromagnetic fluctuations, as in the
work mentioned earlier\cite{afm}. Those diagrams which contain
bubbles mediating an effective interaction between electron propagators
are cancelled, to all orders, by vertex corrections (see Fig. \ref{three}).
Superconductivity arises from the type of diagrams first studied
by Kohn and Luttinger\cite{KL}. The strong anisotropy of the Fermi
surface greatly enhances the Kohn-Luttinger mechanism, with
respect to its effect in an isotropic metal\cite{vanhove}.

Our results support the idea that  d-wave superconductivity and 
antiferromagnetism arise from the same type of interactions.
Antiferromagnetism, however, does not favor the existence
of superconductivity, but competes with it in the same region
of parameter space. Similar physical processes seem to be
responsible for the appearance of anisotropic superconductivity
in systems of coupled repulsive 1D chains\cite{chains}.
The Fermi surface of a single chain is unable to give rise to
this type of superconductivity.  A soon as this limitation is
lifted,  superconductivity occupies a large fraction of the
phase diagram previously dominated by antiferromagnetic
fluctuations.

The results reported above can be extended to fillings away
from the Van Hove singularity, provided that the 
distance of the chemical potential to the singularity is
smaller than the energy scale at which the instability takes place.
The chemical potential tends to be pinned to the singularity
because of the nontrivial RG flow of the chemical potential
itself\cite{vanhove}. Hence, these calculations
can be applied to a finite range of fillings around that
appropriate to the singularity.

In conclusion, we have shown that the $t - t'$ Hubbard model at
the Van Hove singularity exhibits a variety of instabilities at
low energies or temperatures. The existence of these instabilities can 
be derived by RG methods which become
exact at small couplings. We find that antiferromagnetism and 
d-wave superconductivity arise from the same interactions, and
compete with each other in the same region of parameter
space.

\begin{figure}
\caption{Energy contour lines about the Fermi level, with the
Fermi line passing by the saddle points A and B.}
\label{one}
\end{figure}

\begin{figure}
\caption{Different interaction terms arising from the flavor
indexes A and B.}
\label{two}
\end{figure}

\begin{figure}
\caption{Second order diagram renormalizing the different
interactions in the model, with electron lines carrying flavor
index A or B appropriate to each case.}
\label{three}
\end{figure}

\begin{figure}
\caption{Phase diagram in the $(t',U)$ plane. The dotted lines
are contour lines corresponding to the critical frequencies
shown in the figure.}
\label{four}
\end{figure}


\begin{references}


\bibitem{hove}
J. Labb\'e and J. Bok, Europhys. Lett. {\bf 3}, 1225 (1987);
J. Friedel, J. Phys. (Paris) {\bf 48}, 1787 (1987); {\em ibid.}
{\bf 49}, 1435 (1988).

\bibitem{schulz}
H. J. Schulz,  Europhys. Lett. {\bf 4}, 51 (1987).

\bibitem{dzya}
J. E. Dzyaloshinskii,  Pis'ma Zh. Eksp. Teor. Fiz. {\bf 46}, 97 (1987)
[JETP Lett. {\bf 46}, 118 (1987)].

\bibitem{lee}
P. A. Lee and N. Read, Phys. Rev. Lett. {\bf 58}, 2691 (1988).

\bibitem{mark}
R. S. Markiewicz, J. Phys. Condens. Matter {\bf 2}, 665 (1990);
R. S. Markiewicz and B. G. Giessen, Physica (Amsterdam)
{\bf 160C}, 497 (1989).

\bibitem{tsuei}
C. C. Tsuei {\em et al.},
Phys. Rev. Lett. {\bf 65}, 2724 (1990);
P. C. Pattnaik {\em et al.},
Phys. Rev. B {\bf 45}, 5714 (1992).

\bibitem{vanhove}
J. Gonz\'alez, F. Guinea and M. A. H. Vozmediano, Europhys. Lett. {\bf
34}, 711 (1996); Nucl. Phys. B {\bf 485}, 694 (1997).

\bibitem{ioffe}
L. B. Ioffe and A. J. Millis, Phys. Rev. B {\bf 54}, 3645 (1996).

\bibitem{photo}
Z.-X. Shen {et al.}, Science {\bf 267}, 343 (1995), and
references therein.

\bibitem{gofron}
K. Gofron {et al.}, Phys. Rev. Lett. {\bf 73}, 3302 (1994).

\bibitem{zanchi}
D. Zanchi and H. J. Schulz, Phys. Rev. B {\bf 54}, 9509 (1996).

\bibitem{ttp}
P. B\'enard, L. Chen and A. M. Tremblay, Phys. Rev. B {\bf 47},
15217 (1993);
Q. Si, T. Zha, K. Levin and J. P. Lu, {\em ibid.} {\bf 47},
9055 (1993);
A. Nazarenko {\em et al.}, {\em ibid.} {\bf 51}, 8676 (1995);
D. Duffy and A. Moreo, {\em ibid.} {\bf 52}, 15607 (1995);
A. F. Veilleux {\em et al.}, {\em ibid.} {\bf 52}, 16255 (1995).

\bibitem{hirsch}
H. Q. Lin and J. E. Hirsch, Phys. Rev. B {\bf 35}, 3359 (1987).

\bibitem{shankar}
R. Shankar, Rev. Mod. Phys. {\bf 66}, 129 (1994).

\bibitem{pol}
J. Polchinski,
in {\em Proceedings of the 1992 TASI in Elementary Particle
Physics}, J. Harvey and J. Polchinski eds. (World Scientific,
Singapore, 1992).

\bibitem{book}
J. Gonz\'alez, M. A. Mart\'{\i}n-Delgado, G. Sierra and M. A. H.
Vozmediano, {\em Quantum Electron Liquids and High-$T_c$
Superconductivity} (Springer-Verlag, Berlin, 1995).

\bibitem{mc}
T. Husslein {\em et al.}, Phys. Rev. B {\bf 54}, 16179
(1996).

\bibitem{nd}
J. Gonz\'alez and J. V. Alvarez, preprint (cond-mat/9609162), to be
published in Phys. Rev. B.

\bibitem{ferro}
S. Sorella, R. Hlubina and F. Guinea, Phys. Rev. Lett. {\bf 78}, 1343
(1997).

\bibitem{afm}
N. E. Bickers, D. J. Scalapino and S. R. White, Phys. Rev. Lett.
{\bf 62}, 961 (1989);
E. Dagotto, J. Riera and A. P. Young, Phys. Rev. B {\bf 42}, 2347
(1990);
P. Monthoux and D. Pines, Phys. Rev. Lett {\bf 69}, 961 (1992);
E. Dagotto and J. Riera, {\em ibid.} {\bf 70}, 682 (1993);
E. Dagotto, A. Nazarenko and A. Moreo, {\em ibid.} {\bf 74}, 310
(1995).
K. Maki and H. Won, Phys. Rev. Lett. {\bf 72}, 1758 (1994).

\bibitem{KL}
W. Kohn and J. M. Luttinger, Phys. Rev. Lett. {\bf 15}, 524
(1965).



\bibitem{chains}
H.-H. Lin, L. Balents and M. P. A. Fisher, preprint (cond-mat/9703055).








\end{references}
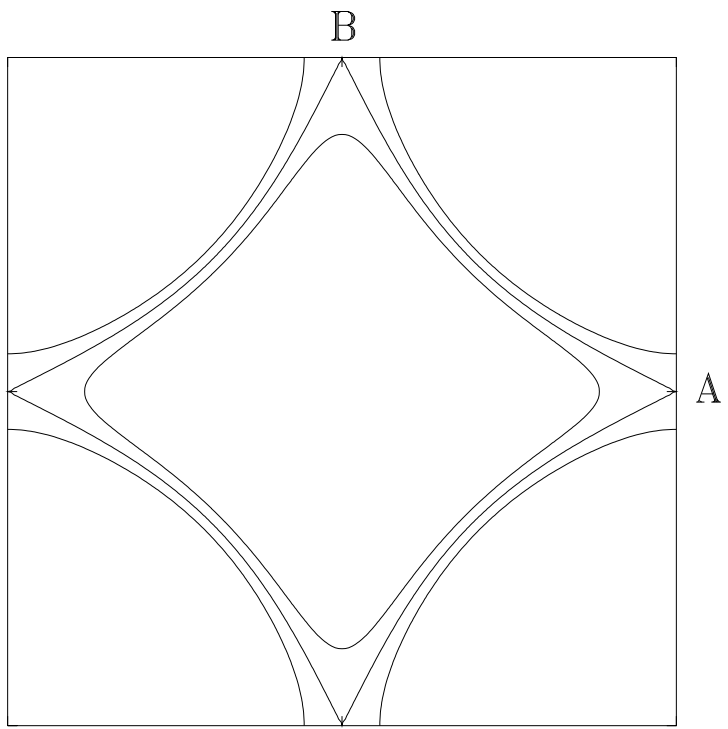
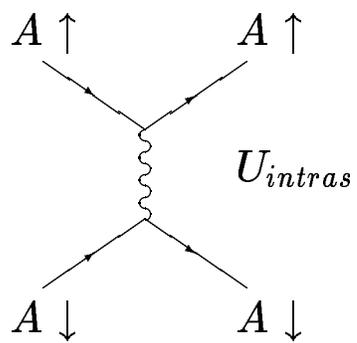
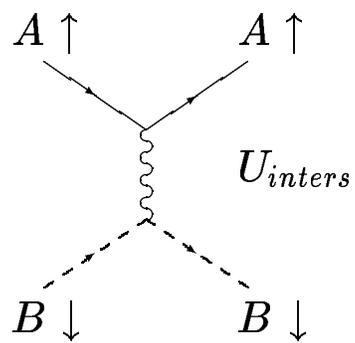
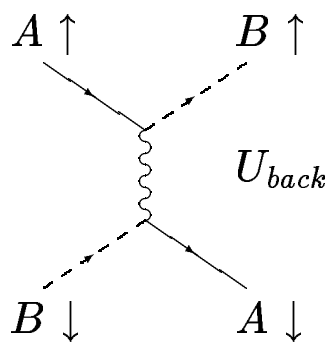
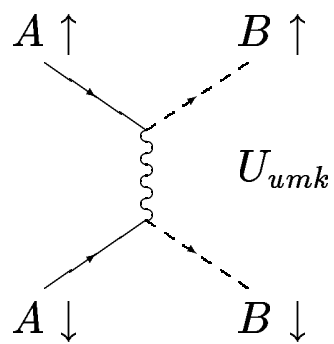
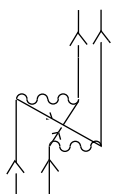
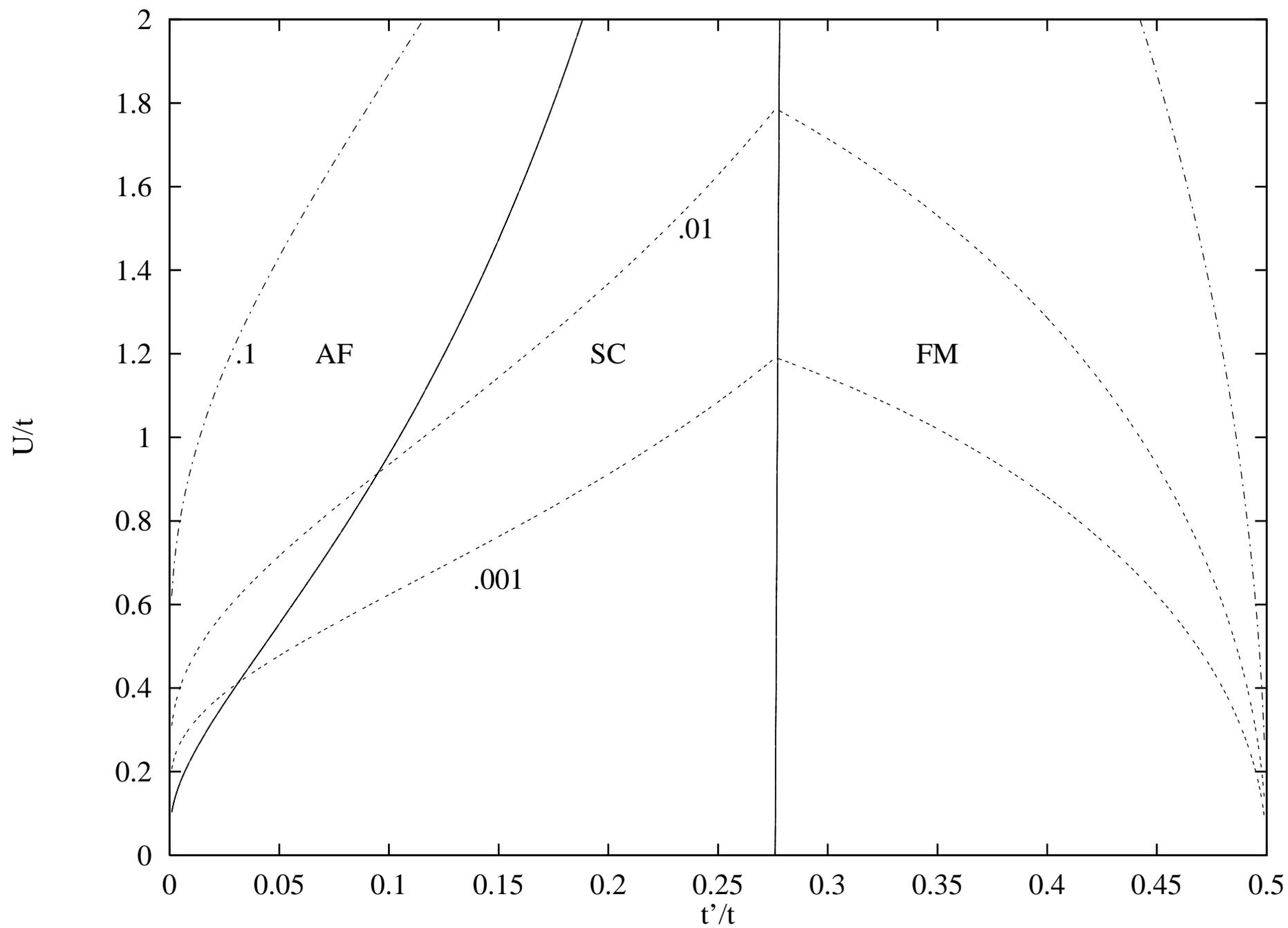
\end{document}